\documentclass[12pt]{article}
\usepackage{amsmath}
\usepackage{jeep}

\input{tcilatex}

\begin{document}

\title{Synchronous Cooperative \ Parrondo's Games}
\author{Zoran Mihailovi\'{c}$^{1}$ and Milan Rajkovi\'{c}$^{2}$ \\
Vin\v{c}a Institute of Nuclear Sciences, \\
P.O. Box 522, 11001 Belgrade, Serbia \\
$^{1}$chesare@vin.bg.ac.yu, $^{2}$milanr@vin.bg.ac.yu}
\maketitle

\begin{abstract}
Inspired by asynchronous cooperative Parrondo's games we introduce two new
types of games in which all players simultaneously play game A or game B or
a combination of these two games. These two types of games differ in the way
a combination of games A and B is played. In the first type of synchronous
games, all players simultaneously play the same game (either A or B), while
in the second type players simultaneously play the game of their choice,
i.e. A or B. We show that for these games, as in the case of asynchronous
games, occurrence of the paradox depends on the number of players. An
analytical result and an algorithm are derived for the probability
distribution of these games.
\end{abstract}

\bigskip

Recently, new types of Parrondo's games have been introduced \cite{Toral},
termed cooperative Parrondo's games that incorporate the feedback through
spatial neighbor dependence. These games are based on the state of player's
nearest neighbors where the state refers to a player either being a winner
or a loser in the previous game. Each of $N$ players, arranged in a circle,
owns a capital $C_{i}(t),i=1,\ldots ,N,$ which evolves by combination of two
games. Game $A$ is the same as in the classical setup \cite{Abot_1},\cite%
{Abot_2},\cite{Abot_3}, namely the probability of winning and losing is $%
p^{(A)}$ and $1-p^{(A)}$ respectively. Game $B$ depends on the state of
neighbors to the left and to the right of the player, i.e. whether they have
won or lost in the previous game. Games $A$ and $B$, when played
individually may be losing or fair, while any kind of periodic or random
alternation of games $A$ and $B$ turns out to be winning. In \cite{Toral},
the evolution of probabilities in games $B$ and $C$ was studied using mean
field type equations while in \cite{Milan}, referred to as paper 1, we have
modeled the games as discrete-time Markov chains and we have derived the
analytic form of the exact probability evolution equations. Inspired by
these games, we introduce here two new types of cooperative games in which
all players play simultaneously either game $A$ or game $B$ or any
combination of these two games and we name these games ``one dimensional
synchronous cooperative games'' in order to make distinction between the
standard cooperative games which we termed ``one dimensional asynchronous
games''. Essentially the difference from the asynchronous case is that at
each turn of the game all players play simultaneously, again depending on
the state of their neighbors, which in turn is the result of the previous
game's outcome. Moreover, a combination of games $A$ and $B$ may be played
in two distinctive ways. First, players may simultaneously play game $A$ or
game $B$ in any predetermined or random order. We denote this game as $A+B$.
Alternatively, each player may at each turn of the game chose randomly
whether to play game $A$ or game $B$, and we denote this game as $A\ast B$.
In both cases the paradoxical result that games $A+B$ or $A\ast B$ may be
winning while each game individually, $A$ or $B$, may be losing, occurs and
it depends on the number of players. Interestingly, for the set of
probabilities introduced in \cite{Toral}, the paradox appears only for $N=3$
and $N=6$. We also introduce a set of probabilities for which a
counterintuitive (paradoxical) result appears irrespective of the number of
players with the exception of $N=4$ and then only for the game of type I ($%
A+B$). As in the case of asynchronous games we derive the probability
transition matrix for games $B$, $A+B$ and $A\ast B.$

As in the case of asynchronous games, the analysis is performed via discrete
time Markov chains (DTMCs) and we first derive the probability transition
matrix for game $B.$ Probabilities of winning in game $B$ depend on the
current state of left and right neighbors, denoted as a pair $h_{k}=(s_{k}-1$
$\ s_{k}+1)$, and with player at position $k$ are given by $p_{0}^{(B)}$
when $(s_{k}-1$ $\ s_{k}+1)$ $=(00)$, $p_{1}^{(B)}$ when $(s_{k}-1$ $\
s_{k}+1)=(01)$, $p_{2}^{(B)}$ when $(s_{k}-1$ $\ s_{k}+1)=(10)$ and $%
p_{3}^{(B)}$ when $(s_{k}-1$ $\ s_{k}+1)$ $=(11)$.

\bigskip

\textbf{Game B}

Let the initial state of the ensemble of players be $i=(i_{1},\ldots ,i_{N})$%
and the final $f=(f_{1},\ldots ,f_{N}).$ Since at each moment of time all
players play simultaneously and therefore change each individual state, the
probability of transition is%
\begin{equation}
T_{fi}^{(B)}=\dprod\nolimits_{k=1}^{N}w(i_{k},f_{k}),  \label{1}
\end{equation}%
where probabilities $w$ depend on whether state $f_{k}$ is $1$ (winning) or $%
0$ (losing), and upon the probability of winning, i.e.%
\begin{equation}
w(i_{k},f_{k})=\left\{ 
\begin{array}{l}
1-p_{\eta _{k}}^{(B)},\text{ }f_{k}=0 \\ 
p_{\eta _{k}}^{(B)},\text{ \ \ \ \ \ \ }f_{k}=1%
\end{array}%
\right.  \label{2}
\end{equation}%
and where $\eta _{k}$ = $(s_{k}-1,$ $\ s_{k}+1)$ = $(i_{k}-1,$ $\ i_{k}+1)$
denotes an ordered pair of $k$-th player's neighbors in the initial state.
Writing $r_{i}=1-p_{i}$ where $i=0,1,2$ or $3$, the probability transition
matrix $\mathcal{T}^{(B)}$\ for $N=3$ is%
\begin{equation}
\mathcal{T}(^{B)}=\left( 
\begin{array}{cccccccc}
r_{0}r_{0}r_{0} & r_{2}r_{1}r_{0} & r_{1}r_{0}r_{2} & r_{3}r_{1}r_{2} & 
r_{0}r_{2}r_{1} & r_{2}r_{3}r_{1} & r_{1}r_{2}r_{3} & r_{3}r_{3}r_{3} \\ 
r_{0}r_{0}p_{0} & r_{2}r_{1}p_{0} & r_{1}r_{0}p_{2} & r_{3}r_{1}p_{2} & 
r_{0}r_{2}p_{1} & r_{2}r_{3}p_{1} & r_{1}r_{2}p_{3} & r_{3}r_{3}p_{3} \\ 
r_{0}p_{0}r_{0} & r_{2}p_{1}r_{0} & r_{1}p_{0}r_{2} & r_{3}p_{1}r_{2} & 
r_{0}p_{2}r_{1} & r_{2}p_{3}r_{1} & r_{1}p_{2}r_{3} & r_{3}p_{3}r_{3} \\ 
r_{0}p_{0}p_{0} & r_{2}p_{1}p_{0} & r_{1}p_{0}p_{2} & r_{3}p_{1}p_{2} & 
r_{0}p_{2}p_{1} & r_{2}p_{3}p_{1} & r_{1}p_{2}p_{3} & r_{3}p_{3}p_{3} \\ 
p_{0}r_{0}r_{0} & p_{2}r_{1}r_{0} & p_{1}r_{0}r_{2} & p_{3}r_{1}r_{2} & 
p_{0}r_{2}r_{1} & p_{2}r_{3}r_{1} & p_{1}r_{2}r_{3} & p_{3}r_{3}r_{3} \\ 
p_{0}r_{0}p_{0} & p_{2}r_{1}p_{0} & p_{1}r_{0}p_{2} & p_{3}r_{1}p_{2} & 
p_{0}r_{2}p_{1} & p_{2}r_{3}p_{1} & p_{1}r_{2}p_{3} & p_{3}r_{3}p_{3} \\ 
p_{0}p_{0}r_{0} & p_{2}p_{1}r_{0} & p_{1}p_{0}r_{2} & p_{3}p_{1}r_{2} & 
p_{0}p_{2}r_{1} & p_{2}p_{3}r_{1} & p_{1}p_{2}r_{3} & p_{3}p_{3}r_{3} \\ 
p_{0}p_{0}p_{0} & p_{2}p_{1}p_{0} & p_{1}p_{0}p_{2} & p_{3}p_{1}p_{2} & 
p_{0}p_{2}p_{1} & p_{2}p_{3}p_{1} & p_{1}p_{2}p_{3} & p_{3}p_{3}p_{3}%
\end{array}%
\right)  \label{3}
\end{equation}%
It may be immediately noticed that this matrix has no zero entries while for
the asynchronous game $B$ it is sparse, i.e. most of its entries are zero.

\bigskip

\textbf{Game A+B (Type I)}

The ensemble of players collectively chooses to play either game $A$ or game 
$B$ so that the probability transition matrix is%
\begin{equation}
\mathcal{T}^{(A+B)}=\gamma \mathcal{T}^{(A)}+(1-\gamma )\mathcal{T}^{(B)},
\label{4}
\end{equation}%
and $\gamma $ represents the relative probability of playing game $A$, where
we assume the value of one half. Matrix $\mathcal{T}^{(A)}$ has a very
simple structure in the unbiased case ( $p^{(A)}=1/2$) when all entries are
equal to $1/8$. In order to illustrate the above expression we calculate the
probability of transition from state $i=(001)$ to state $f=(011)$. The
transition $i\rightarrow f$ occurs when all players play either game $A$ or
game $B$, hence two possibilities are%
\begin{equation}
(001)\rightarrow (011)\left\{ 
\begin{array}{c}
(1-p^{(A)})p^{(A)}p^{(A)}\text{ all players play game }A\text{ i.e. }(AAA)
\\ 
(1-p_{2})p_{1}p_{0}\text{ \ \ \ \ \ \ all players play game }B\text{ i.e. }%
(BBB)%
\end{array}%
\right.   \label{5}
\end{equation}%
Therefore, the probability of transition from state $(001)$ to $(011)$ is%
\begin{equation}
T_{fi}^{(A+B)}=w(001\rightarrow 011)=\gamma \lbrack
(1-p^{(A)})p^{(A)}p^{(A)}]+(1-\gamma )[(1-p_{2})p_{1}p_{0}].  \label{6}
\end{equation}%
Similarly, each entry in the probability transition matrix $\mathcal{T}%
^{(A+B)}$ may be written as the sum according to expression \ref{4}.

\bigskip

\textbf{Game A*B (Type II)}

In this game each player randomly plays either game $A$ or game $B$, thus
individually contributing ``noise'' to the ensemble. In order to shed more
light on the transition process we calculate the transition probability from
state $i=(001)$ to state $f=(011)$. There are $2^{N}=8$ possible ways in
which three player state may change from $i$ to $f$:%
\begin{equation}
(001)\rightarrow (011)\left\{ 
\begin{array}{ll}
(1-p^{(A)})p^{(A)}p^{(A)} & \text{(AAA)} \\ 
(1-p^{(A)})p^{(A)}p_{0} & \text{(AAB)} \\ 
(1-p^{(A)})p_{1}p^{(A)} & \text{(ABA)} \\ 
(1-p^{(A)})p_{1}p_{0} & \text{(ABB)} \\ 
(1-p_{2})p^{(A)}p^{(A)} & \text{(BAA)} \\ 
(1-p_{2})p^{(A)}p_{0} & \text{(BAB)} \\ 
(1-p_{2})p_{1}p^{(A)} & \text{(BBA)} \\ 
(1-p_{2})p_{1}p_{0}\text{ } & \text{(BBB)}%
\end{array}%
\right.   \label{7}
\end{equation}%
Sequences in the second column represent players' choices for each possible
transition. The probability of transition is therefore%
\begin{eqnarray}
T_{fi}^{(A+B)} &=&w(001\rightarrow 011)=\frac{1}{8}%
[(1-p^{(A)})p^{(A)}p^{(A)}+(1-p^{(A)})p^{(A)}p_{0}+  \label{8} \\
&&(1-p^{(A)})p_{1}p^{(A)}+(1-p^{(A)})p_{1}p_{0}+(1-p_{2})p^{(A)}p^{(A)}+ 
\notag \\
&&(1-p_{2})p^{(A)}p_{0}+(1-p_{2})p_{1}p^{(A)}+(1-p_{2})p_{1}p_{0}  \notag \\
&=&[1-(\frac{p^{(A)}+p_{2}}{2})](\frac{p^{(A)}+p_{1}}{2})(\frac{p^{(A)}+p_{0}%
}{2}).  \notag
\end{eqnarray}%
Comparing this expression with the corresponding matrix entry for game $B$,
it may be deduced (and verified by computing probabilities for all
transitions) that the corresponding entries in the probability transition
matrix for game $A\ast B$ may be obtained by replacing each $p_{i}$ $%
(i=0,1,2,3)$ in $\mathcal{T}^{(A+B)}$ with

\begin{equation*}
\frac{1}{2}\left( p^{(A)}+p_{i}\right) .
\end{equation*}%
The equilibrium (stationary) state occurs when the probability distribution
remains invariant under the action of $\mathcal{T}$, that is, $\left| \pi
(t+1)\right\rangle =\mathcal{T}$ $\left| \pi (t\right\rangle =\left| \pi
\right\rangle $, and this probability distribution is determined by solving $%
(\mathbf{1}-\mathcal{T})\pi =0$. For game $A$, for which there is a
probability $p$ for a player to win (alternatively ($1-p$) to lose), the
stationary distribution is easily obtained by setting $%
p_{0}=p_{1}=p_{2}=p_{3}=p$ in (3) or alternatively by associating to each
ensemble state, corresponding probabilities for each player:

\begin{equation}
\pi
^{(A)}=[(1-p)^{3},(1-p)^{2}p,(1-p)^{2}p,(1-p)p^{2},(1-p)^{2}p,(1-p)p^{2},(1-p)p^{2},p^{3}]^{T}
\label{10}
\end{equation}%
For game $B$ the stationary distribution may be obtained analytically using
the probability distribution matrix (3), however the expression is too long
to be presented here. Instead we give the numerical values for the
stationary distributions assuming Torral's values, $p_{0}=1,$ $%
p_{1}=p_{2}=0.16,$ $p_{3}=0.7$:

\begin{equation}
\pi _{1}^{(B)}=[.0570.1906.1906.0849.1906.0849.0849.1161]^{T}  \label{11}
\end{equation}%
\begin{equation}
\pi _{1}^{(A+B)}=[.0972.1495.1495.1033.1495.1033.1033.1444]^{T}  \label{12}
\end{equation}%
\begin{equation}
\pi _{1}^{(A\ast B)}=[.1184.1297.1297.1194.1297.1194.1194.1341]^{T}
\label{13}
\end{equation}%
Interestingly, comparison of expressions \ref{11} and \ref{13} shows that
the stationary probability for the occurrence of state $(000)$ increases
considerably for game $A\ast B$, implying that ``noise'' from game $A$ has
the largest effect on this state when the games are alternating
individually. For the probability values of set II ($p_{0}=0.05,$ $%
p_{1}=p_{2}=0.6,$ $p_{3}=0.8$) we get%
\begin{eqnarray*}
\pi _{2}^{(B)} &=&[.3163.0761.0761.0936.0761.0936.0936.1742]^{T} \\
\pi _{2}^{(A+B)} &=&[.1632.1069.1069.1196.1069.1196.1196.1573]^{T} \\
\pi _{2}^{(A\ast B)} &=&[.1355.1171.1171.1229.1171.1229.12290.1442]^{T}
\end{eqnarray*}%
We now present analytical and numerical results of the model. First we
consider probability values of set I, namely ( $%
p_{0}=1,p_{1}=p_{2}=0.16,p_{3}=0.7$) and $p^{(A)}=0.499$. For this set the
paradox exists only for $N=3$ and $6$ (irrespective whether $A+B$ or $A\ast
B $ game is played) as presented in Fig. 1. Results presented in this figure
were obtained using analytical expression for the evolution of probability
distribution. Moreover, numerical simulations show (Fig.2), that paradox
does not occur for any other number of players up to $1000$. For set II the
paradox exists for all $N$ except for $N=4$, and then only if game $A+B$ is
played. In all other cases the paradox exists as may be noticed in Figs 3
and 4.

In conclusion, one dimensional synchronous games, introduced here and based
on Parrondo's cooperative games exhibit paradoxical outcome characteristic
of classical Parrondo's games, namely that alternation of two losing games
produces a net gain. The novel feature of these games is that all players
play simultaneously either game $A$ or game $B$, as well as a random mixture
of these two games which may be realized in two different ways. Since
winning results from the interaction between different players, these
paradoxical results may be of far-reaching consequences for games of
economic, social and biological importance.

\textbf{Acknowledgement}

This work is financed by Ministry of Science and Technology of Serbia, under
the project OI 1986.

\bigskip

\end{document}